\documentclass[12pt,twoside]{article}
\usepackage{fleqn,espcrc1}

\usepackage{graphicx}
\usepackage[figuresright]{rotating}

\newcommand{\AmS}{{\protect\the\textfont2
  A\kern-.1667em\lower.5ex\hbox{M}\kern-.125emS}}

\title{Probing Quark Matter In Neutron Stars }

\author{M. Prakash\address{Department of Physics \& Astronomy,   
        State University of New York at Stony Brook, \\ 
        Stony Brook, New York-11794-3800, U.S.A.}}
       
\begin{document}

\maketitle

\begin{abstract}
The presence of quark matter in neutron star interiors may have
distinctive signatures in basic observables such as (i) masses and
radii~\cite{LP01}, (ii) surface temperatures versus age~\cite{PPLS00},
(iii) spin-down rates of milli-second pulsars~\cite{GPW97}, and (iv)
neutrino luminosities from future galactic core collapse
supernovae~\cite{Pons01b}.  I highlight recent developments in some of
these areas with a view towards assessing how theory may be confirmed
by $\nu-$signals from future galactic supernovae in detectors like
SuperK, SNO and others under consideration, including UNO~\cite{UNO},
and by multi-wavelength photon observations with new generation
satellites such as the HST, Chandra, and XMM.
\end{abstract}

\section{NEUTRINO SIGNALS }
A proto-neutron star (PNS) is born following the gravitational
collapse of the core of a massive star, in conjunction with a
successful supernova explosion.  During the first tens of seconds of
evolution, nearly all ($\sim$ 99\%) of the remnant's binding energy is
radiated away in neutrinos of all flavors \cite{BL86,PRPLM99,Pons01}.
The $\nu-$luminosities and the evolutionary timescale are controlled
by several factors, such as the total mass of the PNS and the
$\nu-$opacity at supranuclear density, which depends on the
composition and equation of state (EOS).  Collins and Perry
\cite{CP75} noted that the superdense matter in neutron star cores
might consist of weakly interacting quarks rather than of hadrons, due
to the asymptotic freedom of QCD.  The appearance of quarks causes a
softening of the EOS which leads to a reduction of the maximum mass
and radius \cite{LP01}.  In addition, quarks would alter
$\nu-$emissivities and thereby influence the surface temperature of a
neutron star \cite{PPLS00} during the hundreds of thousands or
millions of years that they might remain observable with such
instruments as HST, Chandra, and XMM. 

Many calculations of dense matter predict the appearance of other
kinds of exotic matter in addition to quarks: for example, hyperons or
a Bose (pion, kaon) condensate \cite[and references therein]{Pra97}.
An important question is whether or not $\nu$ observations from a
supernova could reveal the presence of such exotic matter, and further
could unambiguously point to the appearance of quarks.  The detection
of quarks in neutron stars would go a long way toward the delineation
of QCD at finite baryon density which would be complementary to
current Relativistic Heavy Ion Collider experiments, which largely
address the finite temperature, but baryon-poor regime.

An important consequence of the existence of exotic matter in neutron
stars (in whatever form, as long as it contains a negatively charged
component), is that a sufficiently massive PNS becomes metastable
\cite{Pra97,Metas}.  After a delay of up to 100 s, depending upon
which component appears, a metastable PNS collapses into a black hole
\cite{PRPLM99,Pons01}.  Such an event should be
straightforward to observe as an abrupt cessation of $\nu-$flux when
the instability is triggered.

\begin{figure}[ht]
\begin{center}
\includegraphics[scale=0.45]{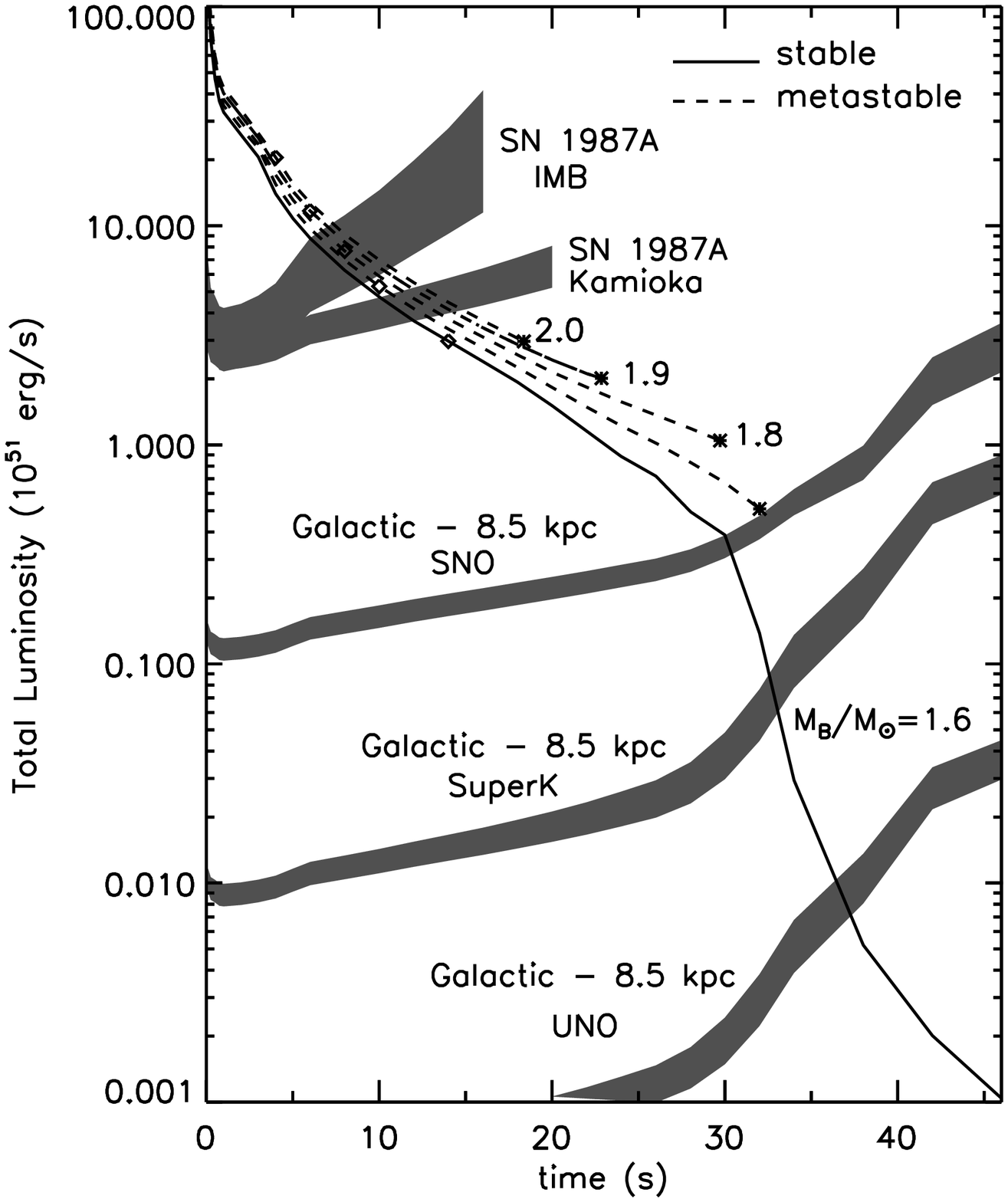} 
\includegraphics[scale=0.475]{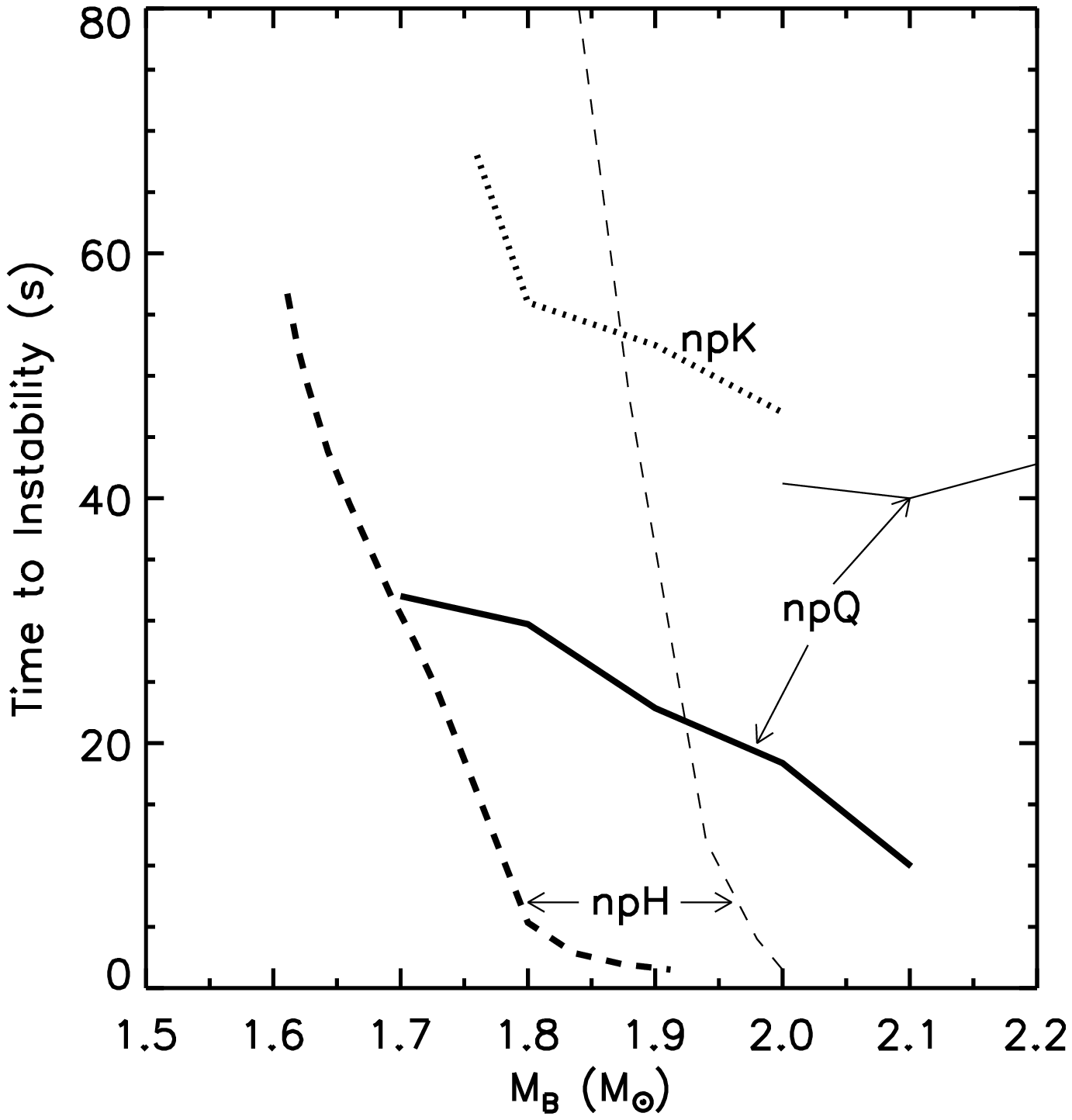}
\end{center}
\vspace*{-0.5in}
\caption{Left panel: The total $\nu-$luminosity for $npQ$ stars of
various baryon masses.  Shaded bands illustrate the limiting
luminosities corresponding to count rates of 0.2 Hz for the indicated
supernovae in some detectors. Right panel: Lifetimes of metastable
stars versus the PNS $M_B$ for various assumed compositions.  Thick
lines denote cases in which the maximum masses of cold, catalyzed
stars are near $M_G\simeq1.45$ M$_\odot$, which minimizes the
metastability lifetimes.  The thin lines for the $npQ$ and $npH$ cases
are for EOSs with larger maximum masses ($M_G=1.85$ and 1.55
M$_\odot$, respectively).}
\label{lum1} 
\end{figure}

In Ref.~\cite{Pons01b} we provide a benchmark calculation with quarks
by solving the general relativistic $\nu-$transport and hydrostatic
equations (see \cite{PRPLM99,Pons01}) with the EOS of \cite{SPL00} and
$\nu$-opacities of~\cite{SPL01} as  microphysical ingredients.  In
the left panel of Fig.~\ref{lum1}, we compare $\nu-$signals
observable with different detectors for stars containing nucleons and
quark matter ($npQ$ stars).  The two upper shaded bands correspond to
estimated SN 1987A (50 kpc distance) detection limits with KII and
IMB, and the lower bands correspond to estimated detection limits set
to a count rate $dN/dt=0.2$ Hz~\cite{Pons01} in SNO, SuperK, and UNO,
for a Galactic supernova (8.5 kpc distance).  It is possible that this
limit is too conservative and could be lowered with identifiable
backgrounds and knowledge of the direction of the signal.  The width
of the bands represents the uncertainty in the $\bar\nu_e$ average
energy due to the flux-limited diffusion
approximation~\cite{PRPLM99,Pons01}.  We conclude that it should be
possible to distinguish between stable and metastable stars, since the
luminosities when metastability is reached are always above
conservative detection limits.

Our quark EOS~\cite{SPL00}, in conjunction with the baryonic EOS we
used, was motivated to maximize the extent of the quark matter phase
in a cold neutron star, and was limited by the necessity of producing
a maximum mass cold star in line with accurate observational
constraints ($M_G=1.444$ M$_\odot$).  Use of an alternative quark EOS
that otherwise produces a larger maximum mass, delays the appearance
of quarks and raises the metastability window to larger stellar
masses~\cite{SPL00}.  Necessarily, this results in an increased
timescale for metastability for a given mass, and a lower
$\nu-$luminosity when metastability occurs.  Fig. \ref{lum1} shows the
relation between time to instability and $M_B$ for the original case
(thick solid curve) and a case (thin solid curve), in which the
maximum gravitational mass of a cold neutron star is about 1.85
M$_\odot$.  For the latter case, the metastability timescales lie in a
narrow range 40--45 s.

In the right panel of Figure~\ref{lum1}, we show the metastability
time-$M_B$ relation found for matter containing hyperons ($npH$,
dashed lines~\cite{PRPLM99}) or matter with kaons ($npK$, dotted
line~\cite{Pons01}) instead of quarks.  All three types of strange
matter are suppressed by trapped neutrinos~\cite{Pra97,SPL00}, but
hyperons always exist in $npH$ matter at finite temperatures and the
transition to quark matter can occur at lower densities than that for
very optimistic kaon cases \cite{Pons01}.  Thus, the metastability
timescales for $npH$ matter can be very short, and those for $npK$
matter are generally larger than for $npQ$ matter.  Note the
relatively steep dependence of the metastability time with $M_B$ for
$npH$ stars, which decreases to very small values near the maximum
mass limit of hot, lepton-rich, stars. The thick $npH$ and $npQ$
lines, as well as the $npK$ line, represent minimum metastability
times for a given $M_B$ as discussed above.  The thin $npQ$ and $npH$
lines are for EOSs with larger cold, catalyzed maximum mass.

Clearly, the observation of a single case of metastability, and the
determination of the metastability time alone, will not necessarily
permit one to distinguish among the various possibilities. Only if the
metastability time is less than 10--15 s, could one decide on this
basis that the star's composition was that of $npH$ matter.  

\newpage
Our conclusions are that (1) the metastability and subsequent collapse
to a black hole of a PNS containing quark matter, or other types of
matter including hyperons or a Bose condensate, are observable in
current and planned $\nu$ detectors, and (2) discriminating among
these compositions may require more than one such observation.  This
highlights the need for breakthroughs in lattice simulations of QCD at
finite baryon density in order to unambiguously determine the EOS of
high density matter.  In the meantime, intriguing possible extensions
of PNS simulations with $npQ$ matter include the consideration of
heterogenoeus structures \cite{CGS00}, quark matter superfluidity
\cite{CR00}, and coherent $\nu-$scattering on droplets \cite{RBP00}.

\section{MULTI-WAVELENGTH PHOTON OBSERVATIONS } 
In Ref.~\cite{PPLS00}, the prospects of detecting baryon and quark
superfluidity from neutron stars during their long-term (up to $10^6$ years) 
cooling epoch was studied.  Our assessment is that, 
from future photon observations of neutron star cooling,  
(1) one could constrain the smaller of the $n-$ or 
$\Lambda-$ pairing gaps and the star's mass, 
(2) deducing the sizes of quark gaps will be difficult,
(3) large $q-$gaps render quarks invisible, and  
(4) vanishing $q-$gaps lead to cooling behaviors which are
indistinguishable from those of $np$ or $npH$ stars. 
 
{\em However, think this titillating thought~!  The observation of a
neutron star older than $10^6$ year and hotter than $\sim 10^{7}~^{\rm
o}{\rm K}$ signals quarks with large gaps in neutron stars~!}

It is a pleasure to thank James M. Lattimer, Dany Page, Jose A. Pons, and
Andrew W. Steiner with whom the work reported here was performed.
This work was supported by the U.S. Department of Energy under
contract number DOE/DE-FG02-88ER-40388.

{}

\end{document}